\documentclass[twocolumn,tighten,twocolappendix]{aastex631}
\usepackage{graphicx}
\usepackage{url}
\usepackage{color}
\usepackage{ragged2e}
\usepackage{hyperref}
\usepackage{url}
\usepackage{amsmath} 
\usepackage{booktabs}
\usepackage{comment}
\usepackage{afterpage}
\usepackage{mathtools}
\usepackage{tabularx}
\usepackage{bold-extra}
\usepackage{xspace}
\usepackage{relsize}
\usepackage{newunicodechar}
\usepackage{lipsum}
\usepackage{multirow}

\usepackage[encapsulated]{CJK}
\usepackage{ucs}
\usepackage[utf8x]{inputenc}

\DeclareGraphicsExtensions{.pdf,.png,.jpeg}

\definecolor{ForestGreen}{rgb}{0.133,0.545,0.133}

\def\rh{r_h}
\def\rp{r_\gamma}
\def\risco{r_{\rm I}}

\def\salehi{Salehi \& Broderick (2023, in preparation)\xspace}

\begin{document}

\title{Shadow Implications: What does measuring the photon ring imply for gravity?}

\author[0000-0002-3351-760X]{Avery E. Broderick}
\affiliation{Perimeter Institute for Theoretical Physics, 31 Caroline Street North, Waterloo, ON, N2L 2Y5, Canada}
\affiliation{Department of Physics and Astronomy, University of Waterloo, 200 University Avenue West, Waterloo, ON, N2L 3G1, Canada}
\affiliation{Waterloo Centre for Astrophysics, University of Waterloo, Waterloo, ON N2L 3G1 Canada}

\author{Kiana Salehi}
\affiliation{Perimeter Institute for Theoretical Physics, 31 Caroline Street North, Waterloo, ON, N2L 2Y5, Canada}
\affiliation{Department of Physics and Astronomy, University of Waterloo, 200 University Avenue West, Waterloo, ON, N2L 3G1, Canada}
\affiliation{Waterloo Centre for Astrophysics, University of Waterloo, Waterloo, ON N2L 3G1 Canada}

\author[0000-0002-3586-6424]{Boris Georgiev}
\affiliation{Perimeter Institute for Theoretical Physics, 31 Caroline Street North, Waterloo, ON, N2L 2Y5, Canada}
\affiliation{Department of Physics and Astronomy, University of Waterloo, 200 University Avenue West, Waterloo, ON, N2L 3G1, Canada}
\affiliation{Waterloo Centre for Astrophysics, University of Waterloo, Waterloo, ON N2L 3G1 Canada}

\begin{abstract}
  With the imaging and characterization of the horizon-scale images of M87* and Sgr A* by the Event Horizon Telescope (EHT), it has become possible to resolve the near-horizon region of astrophysical black holes.  As a result, there has been considerable interest in the implications of the measurement of the shadow size, i.e., the asymptotic photon ring.  We explore the general implications of such a measurement, identifying what is and, more importantly, is not constrained by such measurements, with applications to EHT and future instruments.  We consider a general spherically symmetric metric, which effectively applies for a polar observer (appropriate for M87*) in the slow rotation limit.  We propose a nonperturbative, nonparametric spacetime-domain characterization of shadow size and related measurements that makes explicit the nature and power (or lack thereof) of shadow-size-based constraints, and facilitates comparisons among observations and targets.
\end{abstract}


\section{Introduction}
\label{sec:intro}
With the first images of M87* and Sgr A*, the Event Horizon Telescope has opened a window onto strong gravity in the vicinity of the event horizon of known astrophysical black holes \citep{M87_PaperI,M87_PaperII,M87_PaperIII,M87_PaperIV,M87_PaperV,M87_PaperVI,SgpaperI,SgpaperII,SgpaperIII,SgpaperIV,SgpaperV,SgpaperVI}.  Clearly visible in the images of both sources is the black hole shadow, the locus of rays that intersect the photon sphere (and therefore the event horizon).  This is a purely gravitational feature, and therefore has been the subject of significant interest \citep{Hilbert1917}.  Taken as a measurement of the black hole mass, these observations represent the first time that photon dynamics (the strong lensing of photons) was used to directly weigh a black hole, and confines the mass of the EHT targets into the most compact regions in history.

For both sources, the size of the shadow matches that predicted by general relativity on the basis of stellar dynamical estimates, providing an important confirmation in the strong-gravity regime\footnote{Attempts to measure the shadow shape are unlikely to be fruitful in the near future: M87* is viewed nearly from along the putative spin axis \citep{Spin}, Sgr A* is obscured by a scattering screen and the intervening putative accretion flow \citep{SgpaperVI}.  Thus, we will focus on the shadow size here.} \citep{SgpaperVI,M87_PaperVI}.  However, attempts to quantify the implications for potential deviations from general relativity have made use of either parameterized deviations \citep{Johannsen2013,RZ14,Psaltis2020} or explicit alternative metrics \citep{Kocherlakota2021,SgpaperVI}.  Both of these approaches make strong underlying assumptions that impose strong limits on the interpretation of any results.

Parameterized metric expansions typically suffer from the inherently non-linear nature of general relativity: near the event horizon all terms in the typical expansions become similarly important, obscuring what is a limit and what is an assumption regarding the class of alternative metrics \citep{Psaltis2020,Volkel2020}.  Wherein these assumptions are explicitly avoided by construction \citep[e.g.,][]{RZ14}, the resulting parameter constraints are necessarily strongly correlated --- a natural consequence of a large-dimensional parameter space and a single measurement --- and therefore difficult to interpret practically.  Armed with strong priors, e.g., from gravitational wave experiments or theoretical arguments, this may not be catastrophic.  However, given the novel nature of the EHT images of M87* and Sgr A*, and the extreme mass-scale disparity between EHT and current gravitational wave targets, there is significant value in independent gravitational tests.

Explicit alternative metrics \citep[e.g., those in][]{Kocherlakota2021} provide a physically motivated set of strong priors on the metric deviations by design, and in so doing avoid the arbitrariness associated with a parameterized model.  
However, these constraints are only meaningful within the context of the specific alternative metric under construction, for which the Bayesian prior is unknown and usually assumed to be small.  More importantly, there is no guarantee that neighboring metrics, i.e., ``small'' deviations from the alternative under consideration, are similarly constrained for the same reasons that plague parametric approaches.  Hence, the results from explicit alternatives are typically only interpretable within a narrow context, requiring the onerous reconstruction of images for every metric under consideration.

Here we present an alternative scheme in which to characterize shadow size measurements that restates these in terms of direct measurements of the properties of the metric in an appropriate gauge.  As such, these translate the empirical image-domain measurements performed at infinity to a gravitational domain.  By casting the constraint as a measurement of the metric properties directly, this scheme has two key features:
\begin{enumerate}
    \item It is nonparametric, thus avoiding the complications of strongly correlated parameters while maintaining the general nature of the limit.
    \item It is nonperturbative, and therefore does not require any notion of ``smallness'' and is more naturally applicable in the highly-nonlinear near-horizon regime.
\end{enumerate}
Importantly, by expressing the constraint from shadow size limits in terms of an appropriately specified metric, these are more useful to the gravitational community.  It is no longer necessary to generate full images to compare with the shadow size; rather the computation of the metric components within a convenient gauge at a particular location is sufficient to bring an alternative theory into contact with the EHT and future mm-VLBI constraints.

We will not address the observational problem of measuring the shadow size, which is complicated by the fact that it is only the surrounding luminous plasma that is visible.  \citet{M87_PaperVI} and \citet{SgpaperVI} both calibrate their shadow size measurements with simulated images generated within general relativity or a narrow set of nearby alternatives, rendering interpretation of the size constraints rather more complicated.  More direct methods to extract higher-order images, and therefore infer the diameter of the critical curve that bounds the shadow have been proposed \citep{Spin,Johnson2020}.  However, the measurement particulars and their intrinsic uncertainties are beyond the narrow scope adopted here: what would we learn from such a measurement?  

We will make the simplifying assumptions of spherical symmetry and stationarity.  Even within this restricted class of spacetimes, we are able to elucidate which aspects of the spacetime remain unconstrained by a shadow sizes measurement.  For M87*, which is viewed from nearly the polar axis \citep[see, e.g.,][and references therein]{M87_PaperV,Spin}, spherical symmetry may be relaxed (see, e.g., Salehi \& Broderick 2023, in preparation).  Our conclusions regarding the implications of shadow sizes also hold unchanged in the slow rotation limit for such viewers.

In \autoref{sec:shadows}, we present a formalism for specifying the observed shadow size in spherically symmetric spacetimes, describe our new scheme for characterizing shadow size measurements, discuss the relationship to parameterized and explicit alternative approaches, and discuss the applicability of these results to slowly rotating spacetimes viewed from the spin axis.  We collect the implications of the EHT M87* and Sgr A* campaigns and place these in \autoref{sec:observations}.  These are placed these into the broader context of a variety of non-shadow constraints in \autoref{sec:nonshadow}.  Conclusions are collected in \autoref{sec:conc}.  Unless otherwise specified, we set $G=c=1$.

\section{Shadows from General Spherically Symmetric Spacetimes}
\label{sec:shadows}

\subsection{Definitions and Properties of a General Metric}
We begin with laying out the formalism of generating shadow sizes outside of the limitation of general relativity with a general spherically symmetric metric.  Expressed without loss of generality in areal coordinates, the metric may be written as,
\begin{equation}
    ds^2 = -N(r)^2 dt^2 + \frac{B(r)^2}{N(r)^2} dr^2 + r^2 d\Omega^2.
    \label{eq:metric}
\end{equation}
This metric has two arbitrary real functions of radius, $N(r)$ and $B(r)$, that set the $tt$ and $rr$ components of the metric.  We assume asymptotic flatness, i.e., $\lim_{r\rightarrow\infty} N(r) = 1 - {\mathcal O}(1/r)$ and $\lim_{r\rightarrow\infty} B(r) = 1$.  By construction, we have enforced a metric signature of $(-+++)$ throughout the region of the spacetime accessible to external observers; without loss of generality, we will further assume $N(r)>0$ and $B(r)>0$ everywhere in this region.

This spacetime admits two killing vector fields, and thus two constants of the motion for null geodesics, which we choose to be the energy and angular momentum,
\begin{equation}
    e = p_t = -N^2 \frac{dt}{d\tau}
    ~~\text{and}~~
    \ell = p_\phi = r^2 \frac{d\phi}{d\tau},
\label{eq:consts}
\end{equation}
and due to the spacetime symmetry, all null geodesics are integrable, with the standard properties (e.g., the redshift $1+z=N^{-1}(r)$, etc.).\footnote{Henceforth, we will set $e=1$ and $b=\ell/e$, which due to the assumption of asymptotic flatness is the impact parameter at infinity.}  All black hole spacetimes of this form, by which we mean spacetimes with an event horizon, have $N^2(\rh)=0$ for some horizon radius $\rh$ by definition.  Additionally, all such spacetimes also contain an unstable circular photon orbit (hereafter, abbreviated to simply ``photon orbit''), i.e., a radius at which photons execute a circular orbit about the black hole, located at,
\begin{equation}
  \rp = \frac{N(\rp)}{N'(\rp)},
  \label{eq:rpdef}
\end{equation}
where $N'(r)$ is the radial derivative of $N(r)$.  That an $\rp\ge\rh$ exists that satisfies this condition follows from the assumption of asymptotic flatness and the existence of an event horizon at some finite $\rh$ (see \autoref{app:shadows}).

\subsection{Shadows in Perturbed Spacetimes}
For the innermost shadow, all null geodesics that are outward propagating at $\rp$ will have begun on the horizon (see \autoref{app:shadows}).  Thus, the boundary of the black hole shadow is associated with those null geodesics that are tangent to the photon orbit at $\rp$.  This occurs when the photon angular momentum is equal to a critical value, $b_\gamma=\rp/N(\rp)$.  Identifying $b_\gamma$ with the impact parameter at infinity and using the definition of $\rp$, the shadow radius is
\begin{equation}
    R = \frac{M}{N'(\rp)}.
    \label{eq:Rdef}
\end{equation}
As shown in \salehi and \autoref{app:slowrot}, this continues to hold unchanged for polar observers of slowly spinning black holes (i.e., up to order $a$, where $a$ is the dimensionless black hole spin).  As a result, there is a simple, one-to-one relationship between the observed shadow size and a property of the metric at a specific  location.  From this simple result a number of profound conclusions immediately follow.

\subsection{Characterizing Shadow Size Measurements}
First, because $R$ depends solely on $N'(\rp)$, $N'(\rp)$ is a convenient way in which to characterize the constraints imposed by a shadow size measurement.  While this may appear to be a trivial redefinition given \autoref{eq:Rdef}, the interpretation of $N'(\rp)$ is fundamentally gravitational: it is a direct measurement of spacetime geometry at a dynamically important location for all massless fields, and thus for all electromagnetic and gravitational wave observations.

The fact that shadow-size measurements constrain $N'(\rp)$, and not $N(\rp)$, a point upon which we expound below, suggests a natural framework within which to begin describing near-horizon phenomena generally:
\begin{equation}
    N(\rp),~N'(\rp),~N''(\rp),~\dots
\end{equation}
from which the near-photon-orbit behavior of $N(r)$ can be constructed via Taylor series.  Note that because $R$ depends solely upon $N'(\rp)$, for shadow size measurements this framework is nonperturbative and nonparametric -- measurements of shadow size may be translated into measurements of a (derivative of a) metric coefficient, without making any assumptions about the size and form that any difference from GR might take.  

However, characterizing strong gravity probes in this way is conceptually complicated by the unknown value of $\rp$, i.e., the constrained quantity is the value of $N'(r)$ at the photon orbit, wherever that may be for a particular spacetime.  While at first this may appear an onerous restriction, it is natural in that the location of the photon orbit is gauge invariant (even if the value of $\rp$ may not be).  Explicit examples of applying this constraint will be provided in  \autoref{sec:observations}.

\subsection{$\psi$-$\psi'$ Representation}
An alternative, possibly more familiar representation is in terms of deviations from Schwarzschild, i.e., setting
\begin{equation}
    N^2(r) = 1 - \frac{2M}{r} + 2\psi(r)
    \label{eq:N2psi}
\end{equation}
where in the linearized regime $\psi(r)$ would be the perturbation to the gravitational potential.  When $\psi(\rp)$ is small, $N(\rp)$ and $N'(\rp)$ are simple rearrangements of terms.  However, absent such a guarantee, we have
\begin{equation}
\begin{aligned}
    N(\rp) &= \sqrt{1 - \frac{2M}{\rp} + 2\psi(\rp)}\\
    N'(\rp) &= \left[\frac{M}{\rp^2} + \psi'(\rp)\right] \frac{1}{N(\rp)},
\end{aligned}
\label{eq:NNp2psi}
\end{equation}
with which constraints on $R$ can be converted into a joint constraint on $\psi(\rp)$ and $\psi'(\rp)$:
\begin{equation}
\begin{aligned}
\psi(\rp) &=  \frac{M}{\rp} + \frac{\rp^2}{2R^2} - \frac{1}{2}\\
\psi'(\rp) &= \frac{\rp}{R^2}-\frac{M}{\rp^2}.
\end{aligned}
\label{eq:psipsip}
\end{equation}

Despite appearing more complicated on its face, expressing $N(r)$ in terms of $\psi(r)$ does present one simplification that we will make use of in what follows: if the values of two perturbing potentials and their derivatives match at the $\rp$ for one of them, then $\rp$ is a photon orbit for both and the corresponding shadow sizes are identical.  That is, consider two perturbing potentials, $\psi_1(r)$ and $\psi_2(r)$.  For $\psi_1(r)$, let $\rp$ be the radius of a photon orbit.  Then, if 
\begin{equation}
\begin{aligned}
    \psi_2(\rp)=\psi_1(\rp)
    ~~\text{and}~~
    \psi'_2(\rp)=\psi'_1(\rp),
\end{aligned}    
\end{equation}
$\rp$ will be a photon orbit in the spacetime defined by $\psi_2(r)$.  This follows immediately from the equality of the corresponding $N(\rp)$ and $N'(\rp)$ from \autoref{eq:NNp2psi}, and the use of \autoref{eq:rpdef}.  Moreover, because the shadow size is set by $N'(\rp)$, this is sufficient to guarantee that $R$ is the same for both spacetimes (see \autoref{eq:Rdef}).

As an important specific case, if $\psi(3M)=0$ and $\psi'(3M)=0$, then the photon orbit radius and shadow size are identical to those for Schwarzschild, almost\footnote{We do further assume that there is not an event horizon or second photon orbit at $r>3M$.  These are weak assumptions and do not impact the point being made: a large class spacetimes will have shadow sizes that exactly match those from Schwarzschild.  An explicit example of such a spacetime is provided in \autoref{eq:2pnalt}.} regardless of the form of $\psi(r)$ for $r\ne3M$.

\subsection{Caveats for Known Metric Expansions}
\label{sec:metric_caveats}
As is immediately evident from \autoref{eq:Rdef}, there is no constraint on $N(\rp)$ given a shadow size measurement.  This simple fact has profound consequences for shadow size interpretations: it does not follow that because $R$ matches its general relativistic value that $\psi(r)$ must be small.  This, however, does not mean that shadow size measurements are not constraining; a tight constraint on $R$ does indeed translate into a tight constraint on $N'(\rp)$.  Moreover, the detection of any shadow is a qualitative result that implies $N'(\rp)>0$, eliminating all metrics for which $N(r)$ is decreasing at $\rp$.  These subtleties are often lost in the context of metric expansions, for which $N(r)$ and $N'(r)$ become correlated by construction.  We review some examples here.

\subsubsection{Post-Newtonian Expansion}
The post-Newtonian (PN) formalism employed in \citep{Psaltis2020} and in a restricted manner in \citet{SgpaperVI} presents an expansion of $\psi(r)$ of the form:
\begin{equation}
    \psi(r) = \frac{\kappa_1}{r^2} - \frac{\kappa_2}{r^3} + \frac{\kappa_3}{r^4} - \frac{\kappa_4}{r^5} \dots
    \label{eq:PNpsi}
\end{equation}
where the post-Newtonian coefficients, $\kappa_i$, can be related at large $r$ to terms in the often used parameterized post-Newtonian formalism.  In this way, it is hoped that measurements of $R$ can be related to a broad range of collected tests of general relativity on scales ranging from the laboratory to the cosmos \citep{CMWill2014,Baker2015}  
\begin{figure}
    \centering
    \includegraphics[width=\columnwidth]{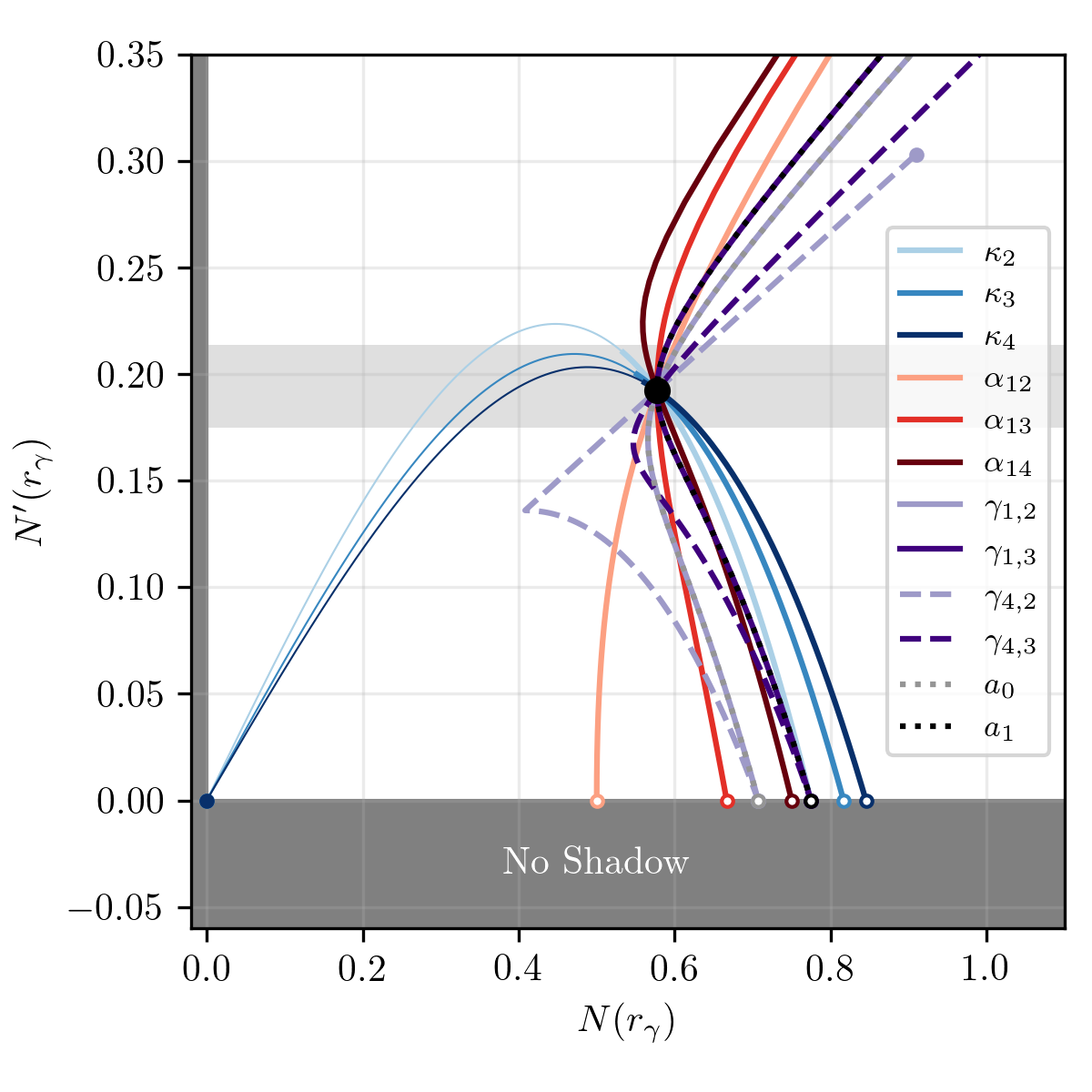}
    \caption{Comparison of allowed range of $N(\rp)$, $N'(\rp)$ for a shadow size measurement with $\sigma_R/R=10\%$ (gray band) and the curves by single-parameter modifications to the metric.  Explicitly, those associated with various PN orders, JP metric orders, MGBK metric orders for $\gamma_1$ and $\gamma_4$, and RZ metric with $a_0$ and $a_1$ varied, holding all other metric expansion parameters fixed at zero (the latter are degenerate with those the MGBK $\gamma_{1,2}$ and $\gamma_{1,3}$ curves, and are otherwise hidden).  Thin lines indicate values for which an event horizon does not exist (affecting only the PN expansion examples); the small open and filled points at which the curves terminate show when the photon orbit becomes infinite and degenerate with the event horizon, respectively.  The large black point indicates the values corresponding to general relativity.}
    \label{fig:2pn}
\end{figure}

As illustrated in \autoref{fig:2pn}, a perturbation defined by any single PN term traces out a curve in the $N(\rp)$-$N'(\rp)$ plane traversed by the associated PN coefficient.  Wherein this curve lies within the band of allowed $N'(\rp)$ given a measurement of $R$, the values of the associated $\kappa_i$ are permissible, appearing to place a constraint on the magnitude of the $\psi(\rp)$.

However, the addition of even a second PN term results in a band that covers the entire physically-relevant quadrant of the $N(\rp)$-$N'(\rp)$ plane.  Thus, it appears that even with only two PN terms, no constraints are possible.  This is, of course, not true: the two PN coefficients are strongly correlated, and it is within the context of that correlation, indicated rather more simply in \autoref{fig:2pn} by the gray band, that the shadow size constraint is present.

Moreover, the measurement of a shadow size does not exclude large deviations beyond $\rp$.  For example, consider the perturbation, 
\begin{equation}
    \psi(r) = \kappa_2 \frac{(3M-r)^2}{r^5},
    \label{eq:2pnalt}
\end{equation}
which is dominated by the 2PN term at $r\gg3M$ with 2PN coefficient $\kappa_2$.  This explicitly satisfies the conditions that $\psi(3M)=0$ and $\psi'(3M)=0$, and thus has $\rp=3M$ and $R=\sqrt{27}M$, identical to those from Schwarzschild.  This is true for any value of $\kappa_2$. In this sense, there is no meaningful limit on $\kappa_2$ from any measurement of the dynamics of massless fields near the photon orbit without additional, typically strong, assumptions about the spacetime geometry.

\subsubsection{Other Metric Expansions}
For completeness, in \autoref{fig:2pn}, we also show the paths traced out by the various other metric expansions considered in Section 5.1 of \cite{SgpaperVI}.  These expansions include the spherically-symmetric restriction on the metrics proposed by \citet[hereafter JP]{Johannsen2013}, MGBK \citet[hereafter MGBK]{Vigeland2011}, and \citet[hereafter RZ]{RZ14}, as described in \citet{SgpaperVI}.  Each exhibits a similar qualitative behavior to the post-Newtonian expansions: the range of the inferred limits on $N(\rp)$ are solely due to the priors imposed by the underlying expansion themselves.  Quantitative differences are present, further highlighting the impact of these priors.

\subsection{Comparison to Explicit Alternatives}
Alternative metrics to Schwarzschild, e.g., Reissner-Nordstr\"om and those associated with alternative gravity theories, present a similar story as those associated with metric expansions.  The details of the metric perturbation appear to induce a limit on $N(\rp)$ through model-induced correlations between $N(\rp)$ and $N'(\rp)$.  However, alternative metrics differ in an important conceptual way: the correlations are a consequence of the physical prior that the metric of interest applies and is not an arbitrary truncation of an otherwise infinite series of terms.  In this sense, the constraints are meaningful within the narrow context of the alternative metric.

\begin{figure}
    \centering
    \includegraphics[width=\columnwidth]{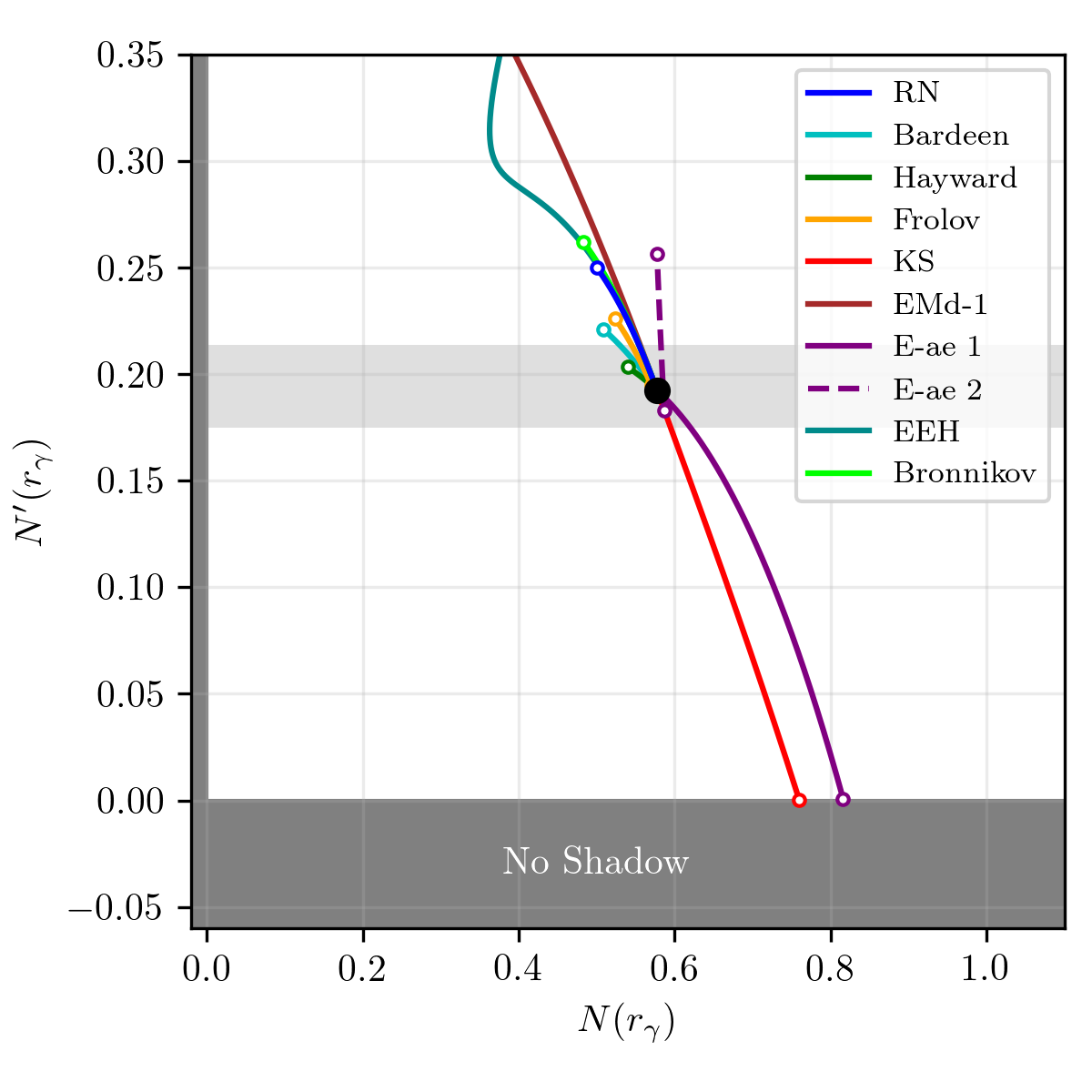}
    \caption{Comparison of allowed range of $N(\rp)$, $N'(\rp)$ for a shadow size measurement with $\sigma_R/R=10\%$ (gray band) and the curves associated with the various spherically symmetric alternative spacetimes considered in \citet{Kocherlakota2020}.  The small open points at which the curves terminate show when the charge reaches an intrinsic limit, e.g., the solution no longer has an event horizon. The large black point indicates the values corresponding to general relativity.}
    \label{fig:alts}
\end{figure}

In \autoref{fig:alts} we show the regions of the $N(\rp)$-$N'(\rp)$ plane spanned by a sample of alternative black hole metrics.  Following \citet{SgpaperVI}, we focus attention on a subset of representative spherically-symmetric alternatives, though we expand this list to the twelve listed in Table 1 of \citet{Kocherlakota2020}. We do not make any representation that these twelve are complete, but rather only that they are illustrative.\footnote{Where a metric depends on more than one parameter, we evaluate it at a fixed value of all but one.}  We refer the reader \citet{SgpaperVI} and \citet{Kocherlakota2020} for details on the metrics themselves and their underlying assumptions.  

Two things are immediately evident upon comparison with the parameterized metric expansions.  First, the imposition of physical constraints on the metric itself typically limits the region in the $N(\rp)$-$N'(\rp)$ plane spanned by alternative metrics significantly.  In this sense, the parameterized metric expansions are more agnostic, covering a wider variety of potential deviations from general relativity.  However, this is also a consequence in the difference in interpretation: where the metric expansions need to be sensible only locally, alternative spacetimes must be globally well-behaved.  

Second, the general direction in the $N(\rp)$-$N'(\rp)$ spanned by the alternative metrics differs from those for any single-parameter exploration in the metric parameterizations in \autoref{fig:2pn}.  Of course, upon permitting more than one parameter to vary in the metric expansions it is possible to mimic the alternative spacetimes \citep[e.g., see Section IV of][]{Kocherlakota2020}.  However, this illustrates the difficulties faced by single-parameter characterizations of the shadow-size constraints.

\section{Implications of known shadow sizes}
\label{sec:observations}
We now review the implications of measurements of the shadow sizes arising from EHT observations of M87* and Sgr A*.  While some uncertainty regarding the methodology of such measurements may persist, we take these at face value here, and assess the implications for deviations from gravity using the $N(\rp)$-$N'(\rp)$ formalism and more traditional metric expansions and alternatives.  

\begin{figure*}
    \centering
    \includegraphics[width=\columnwidth]{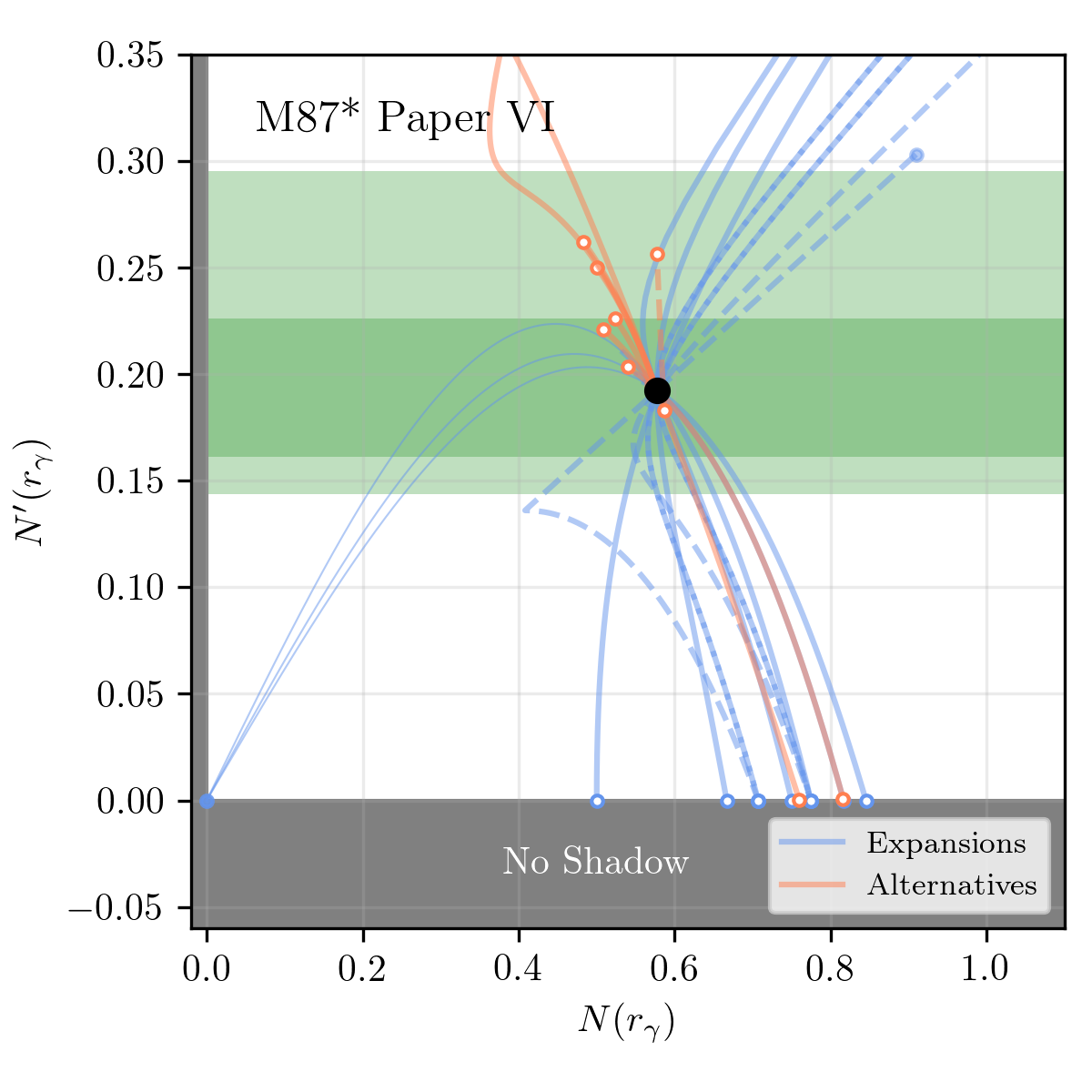}
    \includegraphics[width=\columnwidth]{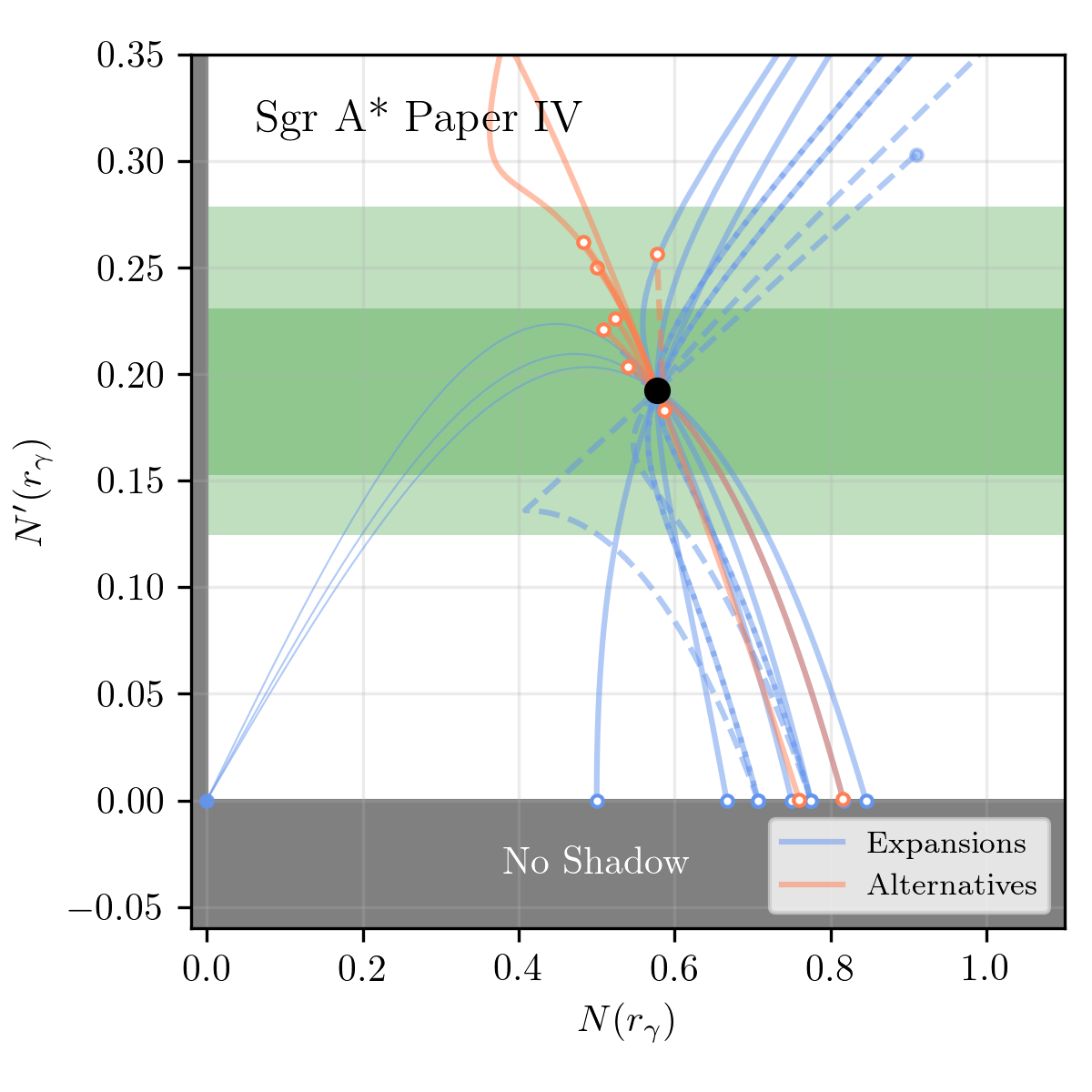}
    \caption{Representative constraints implied by the shadow size constraints for M87* (left) and Sgr A* (right) published by the EHT.  Illustrative curves associated with the metric expansions in \autoref{fig:2pn} (blue) and alternative spacetimes in \autoref{fig:alts} (orange) are shown.  See \autoref{fig:2pn} for a description of line types.  In both panels the relevant EHT measurements at $1\sigma$ and $2\sigma$ are shown by the green bands.}
    \label{fig:ehtlimits}
\end{figure*}

\begin{deluxetable*}{ccccccc}
\caption{Direct and Apparent Implications of EHT Shadow Sizes for Various Metrics}
\label{tab:ehtlimits}
\tablehead{
&
\multirow{2}{*}{
\raisebox{-0.30cm}{
\begin{minipage}{2cm}
\centering
Sole Varied\\
Parameter
\end{minipage}}}
&
\multicolumn{2}{c}{M87* 2017}
&
&
\multicolumn{2}{c}{Sgr A* 2017}\\
\cline{3-4} \cline{6-7}
\colhead{Basis/Metric}
&
&
\colhead{Paper VI}
&
\colhead{$\theta_{n=1}$}
&
&
\colhead{Paper VI}
&
\colhead{Paper IV}
}
\startdata
\hline
\multirow{2}{*}{This paper}  & $\sqrt{3}N(\rp)$   & \nodata        & \nodata       && \nodata       & \nodata\\
     & $\sqrt{27}N'(\rp)$ & $0.95_{-0.12}^{+0.22}$ & $0.89_{-0.07}^{+0.16}$ && $1.04_{-0.10}^{+0.12}$ & $1.02_{-0.23}^{+0.17}$ \\
\hline
\multirow{3}{*}{PN}   & $\kappa_2$     & $[-2.7,0.7]$   & $[-3.2,0.7]$  && $[-0.5,0.7]$  & $[-4.2,0.7]$\\
     & $\kappa_3$     & $[-10.9,1.1]$  & $[-13.0,1.0]$ && $[-1.7,1.2]$  & $[-17.7,1.0]$\\
     & $\kappa_4$     & $[-44.1,1.8]$  & $[-54.1,1.8]$ && $[-5.5,2.2]$  & $[-77.6,1.8]$\\
\hline
\multirow{3}{*}{JP}   & $\alpha_{12}$  & $[-1.2,2.0]$   & $[-0.4,2.2]$  && $[-1.1,0.5]$  & $[-1.3,2.8]$\\
     & $\alpha_{13}$  & $[-3.2,7.0]$   & $[-1.1,8.0]$  && $[-3.1,1.5]$  & $[-3.5,10.2]$\\
     & $\alpha_{14}$  & $[-8.2,26.0]$  & $[-3.2,30.5]$ && $[-7.8,4.6]$  & $[-8.7,40.5]$\\
\hline
\multirow{4}{*}{MGBK} & $\gamma_{1,2}$ & $[-3.2,3.2]$   & $[-0.8,3.5]$  && $[-2.9,0.9]$  & $[-3.6,4.2]$\\
     & $\gamma_{1,3}$ & $[-8.6,11.5]$  & $[-2.5,13.1]$ && $[-8.0,2.7]$  & $[-9.8,16.2]$\\
     & $\gamma_{4,2}$ & $[-5.2,4.0]$   & $[-1.3,4.4]$  && $[-4.8,1.3]$  & $[-5.9,5.0]$\\
     & $\gamma_{4,3}$ & $[-15.2,13.0]$ & $[-3.9,14.3]$ && $[-14.0,3.8]$ & $[-17.5,16.8]$\\
\hline
\multirow{2}{*}{RZ}   & $a_0$          & $[-0.8,0.8]$   & $[-0.9,0.2]$  && $[-0.2,0.7]$  & $[-1.0,0.9]$\\
     & $a_1$          & $[-1.4,1.1]$   & $[-1.6,0.3]$  && $[-0.3,1.0]$  & $[-2.0,1.2]$\\
\hline
Reissner-Nordstr\"om   & $0<\bar{q}\le 1$           & $<0.86$ & $<0.51$ && $<0.84$ & $<0.90$\\
Bardeen                & $\bar{q}_m\le\sqrt{16/27}$ & \nodata & $<0.50$ && \nodata & \nodata\\
Hayward                & $\bar{l}\le1.06$           & \nodata & $<1.01$ && \nodata & \nodata\\
Frolov ($\bar{l}=0.4$) & $\bar{q}\le0.79$           & \nodata & $<0.42$ && $<0.77$ & \nodata\\
Kazakov-Solodhukin     & $\bar{a}$                  & $<1.65$ & $<1.75$ && $<0.79$ & $<1.95$\\
EMd-1                  & $\bar{q}\le\sqrt{2}$       & $<0.90$ & $<0.51$ && $<0.87$ & $<0.94$\\
E ae-1                 & $0\le\bar{c}_{13}<1$       & $<0.93$ & $<0.94$ && $<0.66$ & $<0.95$\\
E ae-2 ($\bar{c}_{13}=0.25$) & $0\le\bar{c}_{14}\le2\bar{c}_{13}$ & $<0.33$ & $<0.16$ && $<0.31$ & $<0.36$\\
CFM A                  & $\beta<1$                  & \nodata & \nodata && \nodata & \nodata\\
CFM B                  & $1<\beta<5/4$              & \nodata & \nodata && \nodata & \nodata\\
EEH ($\bar{\alpha}=0.25$) & $0<\bar{q}_m$           & $<0.86$ & $<0.51$ && $<0.84$ & $<0.90$\\
Bronnikov              & $0\le\bar{c}_{13}<1$       & $<0.86$ & $<0.51$ && $<0.84$ & $<0.90$
\enddata
\tablecomments{For all metrics, only the parameter listed is varied.  Varying multiple parameters typically results in no parameter constraint.  Missing entries correspond to no constraints.  See \autoref{sec:shadow_measurement} for how the various shadow size estimates are produced.}
\end{deluxetable*}

\subsection{Shadow Size Estimates}
\label{sec:shadow_measurement}
We consider four EHT shadow size measurements arising from the 2017 observing campaign, two each for M87* and Sgr A*, differing in the particulars in how they are produced.  All involve two underlying measurements, that of the shadow size with the EHT and a comparison mass measurement.  

\subsubsection{2017 M87* Paper VI}
First, for M87*, we adopt the angular size of the gravitational radius, i.e., $\theta_g=GM/c^2 D$ where $M$ and $D$ are the mass of and distance to M87*, reported in \citet{M87_PaperVI} of $3.8\pm0.4\,\mu{\rm as}$.  Stellar dynamics measurements by \citet{Gebhardt2011} produce a corresponding estimate of $\theta_{\rm dyn}=3.62^{+0.60}_{-0.34}\,\mu{\rm as}$ \citep{M87_PaperVI}.  Taking the latter to define $M$, the former gives an estimated Shadow radius of
\begin{equation}
    R/M = \sqrt{27} \frac{\theta_g}{\theta_{\rm dyn}}
    = \sqrt{27} \left( 1.05_{-0.20}^{+0.15} \right).
\end{equation}
This estimate presumes that the shadow size is indeed related to the EHT-measured size by the canonical $\sqrt{27}$, which may differ due to spin and/or the assumed astrophysics of the emitting region \citep{Gralla2019,Blandford2022}.  The latter concern is amplified by the fact that the mass measurement presented in \citet{M87_PaperVI} calibrates the relationship between the bright ring and black hole mass using simulations that assume general relativity.  Nevertheless, it forms the basis for the general relativity tests reported in \citep{M87_PaperVI} and \citet{Psaltis2020}, and thus we include it here.  We refer to this shadow size measurement as the M87* 2017 Paper VI estimate in \autoref{tab:ehtlimits}.

\subsubsection{2017 M87* $\theta_{n=1}$}
Second, again for M87*, we use the size of the $n=1$ photon ring $\theta_{n=1}=21.74\pm0.10\,\mu{\rm as}$, generated via the secondary image of the emission region, reported in \citet{Broderick2022}.  This is calibrated using numerical simulations to the shadow size, which finds a shift between the two of $\Delta\theta=0.56\pm0.32\,\mu{\rm as}$, resulting in a spin-zero estimate of the shadow radius of $21.09\pm0.33\,\mu{\rm as}$.  Again using the stellar dynamics estimate of $M$, this gives a shadow radius of
\begin{equation}
    R/M = \sqrt{27} \left( 1.12_{-0.17}^{+0.10} \right).
\end{equation}
This estimate is less dependent on the underlying astrophysics, though does rely upon the identification of the ring-like structure with the $n=1$ photon ring.  It is, however, again reliant upon the numerical simulations used to relate the $n=1$ photon ring to the edge of the shadow.  We refer to this shadow size measurement as the M87* 2017 $\theta_{n=1}$ estimate in \autoref{tab:ehtlimits}.

\subsubsection{2017 Sgr A* Paper VI}
Third, for Sgr A*, we make use of the estimate reported in \citet{SgpaperVI} via the $\delta$:
\begin{equation}
    R/M = \sqrt{27} \left( 1 + \delta \right)
    = \sqrt{27} \left( 0.96 \pm 0.10 \right).
\end{equation}
This estimate depends on the specific calibration procedure employed in \citet{SgpaperVI}, which differs from that used for M87* in some important respects: the calibration is performed directly to the shadow size, included in the calibration library are a handful of non-Kerr spacetimes.  However, it also makes astrophysical assumptions regarding the emission region, and optimistic assumptions regarding the variability.  As a result, we consider this to be an optimistic constraint on the shadow size.  We refer to this shadow size measurement as the Sgr A* 2017 Paper VI estimate in \autoref{tab:ehtlimits}.

\newpage
\subsubsection{2017 Sgr A* Paper IV}
Finally, for Sgr A*, we employ a similar procedure as that employed by M87* based upon the mass estimate in \citet{SgpaperIV}, $\theta_g=4.8_{-0.7}^{+1.4}\,\mu{\rm as}$, and the stellar dynamics measurements by \citet{Do2019} and \citet{Gravity2022} as collated in \citep{SgpaperVI}, $\theta_{\rm dyn}=4.92\pm0.3\,\mu{\rm as}$.  The resulting estimate for the shadow size is
\begin{equation}
    R/M = \sqrt{27}  \frac{\theta_g}{\theta_{\rm dyn}}
    = \sqrt{27} \left( 0.98_{-0.14}^{+0.28} \right).
\end{equation}
This is significantly more conservative than the estimate from \citep{SgpaperVI}, with roughly twice the uncertainty.  We refer to this shadow size measurement as the Sgr A* 2017 Paper IV estimate in \autoref{tab:ehtlimits}.

\subsection{Metric Expansions}
\label{sec:exps}
The left panels of \autoref{fig:ehtlimits} shows the implications for parameterized metric perturbations of the 2017 M87* Paper VI and 2017 Sgr A* Paper IV shadow size estimates.  As in \autoref{fig:2pn}, each metric parameterization imposes a strong prior within the $N(\rp)$-$N'(\rp)$ plane, inducing an apparent constraint on $N(\rp)$ given the shadow measurement's direct constraint on $N'(\rp)$.  

The $1\sigma$ single-parameter limits, when all other perturbations are fixed at zero, are collected in \autoref{tab:ehtlimits} for all four shadow size estimates.  These ranges are obtained simply by inspecting the range of parameter values for which the curves in \autoref{fig:ehtlimits} remain in the $1\sigma$ bands.  Note in particular that the limits from the 2017 Sgr A* Paper VI shadow size estimates quantitatively match those in \citet{SgpaperVI}.\footnote{A solitary exception is the constraint on $\gamma_{1,2}$ of the MGBK expansion.  However, the values in \autoref{tab:ehtlimits} quantitatively match those inferred from Fig.~17 of \citet{SgpaperVI}, and thus we attribute the mismatch to a typographical error in Table~3 of \cite{SgpaperVI}.}  However, all of these limits should be interpreted with significant care for the reasons described in \autoref{sec:metric_caveats}.  In particular, the constraint on the magnitude of the perturbation, at $\rp$ or otherwise, is illusory.

\subsection{Alternative Spacetimes}
The implications for the alternative metrics in \citet{Kocherlakota2020} are shown in the right panes of \autoref{fig:ehtlimits} for the 2017 M87* Paper VI and 2017 Sgr A* Paper IV shadow size estimates.  The $1\sigma$ limits on the alternative metrics' charges are collected in \autoref{tab:ehtlimits} for all four shadow size estimates.  Where appropriate, these agree with those reported in \citet{SgpaperVI} and \citet{Kocherlakota2021}.

Generally, the global constraints on the alternative metrics, e.g., the necessity of an event horizon, limits the range of $N'(\rp)$ permitted substantially.  While not universal, this does impose a typical scale on the shadow size measurements that will be informative, roughly requiring measurement precisions of a few percent.

As with the parameterized metric expansions, limits on $N(\rp)$ are inferred from the form of the particular alternative metric under consideration.  Therefore, further empirical progress requires additional observables that constrain quantities other than $N'(\rp)$.

\section{Beyond Shadow Sizes}
\label{sec:nonshadow}
A variety of astrophysical probes of the strong-gravity regime have either already become possible, or will be possible in next decade.  These are frequently relevant for ostensibly stationary spacetimes, e.g., presumably Kerr black holes, and we will consider that case.  Here, we explore how they may also be incorporated into measurements of $N(\rp)$, $N'(\rp)$, etc., and what additional underlying assumptions may be necessary.

\subsection{Light Echos}
Echoes associated with time delays between the direct emission and higher order images have been proposed as an alternate probe of the spacetime \citep{Moriyama2019,Hadar_2021}.  These time delays are a result of the additional path length of the photon trajectories associated with the higher order images as the photons orbit about the black hole prior to streaming toward the observer at infinity.  Thus, the typical timescale is the orbital time at the photon orbit, as measured by a distant observer.

Naively, one might imagine that this period is $2\pi\rp$, and thus a direct measure of the radius of the photon orbit.  However, due to the gravitational redshift, in practice the measurable orbital period is
\begin{equation}
    P = \frac{2\pi\rp^2}{b_\gamma N^2(\rp)} = 2\pi R.
\end{equation}
Hence, light echos and other orbital phenomena at the photon orbit provide degenerate information to that contained within the measurement of the shadow size.

\subsection{Doppler Effects}
The orbiting material responsible for the emission will be Doppler beamed and shifted due to the bulk motions.  This effect is responsible for the typically asymmetric images appearing in simulated accretion flows and observed in M87*.  While the flux is dependent on the density and magnetization of the emitting material, under the assumption of an approximately axisymmetric emission region (at least on average), the ratio of fluxes on the approaching and receding sides will depend solely on the dynamics of the orbiting plasma and spectral index of the synchrotron emission.  

Using the Lorentz invariance of $I_\nu/\nu^3$, where $I_\nu$ is the observed intensity at a location on the sky at frequency $\nu$, the flux ratio is, 
\begin{equation}
    {\mathcal F} = \left(\frac{\nu_r}{\nu_a}\right)^{\alpha+3},
\end{equation}
where we have assumed the emitted intensity is $\propto \nu^{-\alpha}$, and $\nu_a$ and $\nu_r$ are the photon frequencies in the frame of the emitting plasma that is approaching or receding, respectively.  These frequencies are related to the photon four-vector, $k^\mu$, and plasma four-velocity, $u^\mu$, at the point of emission (which we will assume is located at $\rp$) for a given location on the observing screen ($R$), and are therefore sensitive to astrophysical assumptions made about the accretion flow and emission mechanism.

For concreteness, if we presume that the emission arises from a thin, modestly sub-Keplerian accretion disk, i.e., the disk angular velocity as measured by a distant observer is
\begin{equation}
    \Omega^2 = \kappa^2 \frac{N(r) N'(r)}{r},
    \label{eq:Omega2}
\end{equation}
which at the photon orbit reduces to $\kappa^2/R^2$, where the sub-Keplerian factor is $\kappa\sim0.9$ in GRMHD simulations \citep{Narayan_2012,Porth_2019}.  Near the edge of the shadow, for an observing inclination of $i$ the observed frequencies for the approaching and receding sides are
\begin{equation}
    \nu_{a,r}
    =
    \nu_{\rm obs} u^t \left( 1 \mp \kappa \sin i \right)
\end{equation}
which, up to the multiplicative factor $\nu_{\rm obs} u^t$, is independent of $N(\rp)$ and $N'(\rp)$.  As a consequence, under the above astrophysical assumptions, measurements of $\mathcal{F}$ does not provide an additional constraint on the underlying spacetime.\footnote{Of course, it does provide insight into $\kappa$ and $i$, both of which are of considerable astrophysical interest!}

A substantially different astrophysical picture of the near-horizon emission region may result in different spacetime dependencies.  However, in that case, the gravitational measurements are fundamentally subject to the currently large astrophysical uncertainties.  While not insurmountable with the inclusion of additional observations (e.g., multi-wavelength views, variability studies, spectral energy distributions, etc.) a full discussion of how to perform joint astrophysical/gravitational analyses is beyond the scope of this paper.

\subsection{Redshift Measurements}
The gravitational redshift from emission near the photon orbit is dependent solely on $N(\rp)$:
\begin{equation}
    1+z = 1/N(\rp).
\end{equation}
As with the approach/receding flux ratio discussed above, this is critically dependent upon the dynamical state of the emission region, and thus mixes the astrophysical and gravitational uncertainties.

Relativistically broadened iron fluorescence lines (Fe K$\alpha$) have been detected in a number of quasars \citep{Perez2010,Tanaka1995}. These X-ray lines are the combined emission from a number of radii, and are impacted by the X-ray corona, the rapid orbital motions, and the gravitational redshift, which generates a characteristic broad ``red wing'' that extends to lower X-ray energies.  Typically, this emission is assumed to arise outside of the innermost stable circular orbit \citep[ISCO;][]{Reynold1997}, which for Schwarzschild lies well beyond $\rp$.  We discuss the uncertain relationship between the ISCO and $\rp$ in \autoref{sec:emri}, and here note only that this makes direct comparisons with the redshifts from these systems and $N(\rp)$ challenging.

For M87* and Sgr A*, the redshift has a further problem: the need for spectral features from which to estimate a redshift.  The observed spectral energy distributions of M87* and Sgr A* exhibit no features beyond the broad-band optically thick to thin transition near EHT wavelengths.  The nearly virialized accretion flows believed to be present in these sources is expected to reach temperatures above $10^{10}\,{\rm K}$ near the horizon.  At these temperatures, were the accreting gas in local thermodynamic equilibrium (LTE), all atomic species are photoionized, and most nuclei are photodissociated.  However, the low densities within the accretion flows, which are that are directly responsible for their radiative inefficiency, also imply that the gas will typically not be in LTE, opening the possibility that tightly bound atomic species (e.g., Fe) may persist in the near-horizon region.  In addition, nuclear gamma-ray lines will be present. Finding and measuring either of these would require next generation instruments.

\subsection{Multiple Photon Rings}
The shadow is bounded by an infinite sequence of higher-order images of the accretion flow, often referred to as ``photon rings''.  Each higher-order photon ring is the result of an additional half-orbit executed by the null geodesic prior to streaming toward a distant observer.  Necessarily, each ring is distinctly located in the image (no position in the image can contribute to more than one null geodesic, and, therefore, more than one image).  These are purely geometric features, depending only upon the strong lensing within the spacetime, and therefore present a natural probe of general relativity \citep{Spin}.

The ability to separate and extract individual ring-like structures within images has recently been developed \citep[e.g.,][]{Themaging,Broderick2022} and future experiments may be able to resolve multiple such rings \citep{Johnson2020,Spin}.  Thus, measuring the size of multiple photon rings presents a natural and practical extension to the notion of measuring the shadow size, and is necessarily probing the region near the photon orbit.

In \autoref{app:photonrings} we estimate the radii of high-order photon rings relative to that of the shadow, finding,
\begin{equation}
    R_n - R \approx f e^{-\gamma n},
\end{equation}
where $f$ is a function of $\rp$, $N(\rp)$, $N'(\rp)$, and $N''(\rp)$, and
\begin{equation}
    \gamma = \pi \frac{N^{3/2}(\rp)}{N'(\rp)} 
    \left[ -\frac{N''(\rp)}{B^2(\rp)} \right]^{1/2}
\end{equation}
is the Lyapunov exponent that defines the self-similar ring structure.  Note that in addition to $N(\rp)$ and $N'(\rp)$, $\gamma$ depends on $N''(\rp)$ and $B(\rp)$ through the combination $N''(\rp)/B^2(\rp)$. This is a consequence of the fact that the finite order photon ring radii are dictated by the dynamics of photons very nearby, but outside of the photon orbit.  Because $\gamma$ also describes the rate at which the radii of these trajectories grow, it naturally depends on the second derivative of the effective potential, $N''(\rp)$, and the notion of radial distance, $B(\rp)$, at $\rp$.

The absolute normalization must ultimately be computed numerically and may differ among spacetimes.  However, the relative sizes are fully fixed by $\gamma$, and thus it is possible to measure $\gamma$ directly with a shadow size and two photon ring radii:
\begin{equation}
    \gamma = \ln\left(\frac{R_{n}-R}{R_{n+1}-R}\right).
\end{equation}
Alternatively, measuring three photon ring radii permits removing $R$ altogether,
\begin{equation}
    \gamma = \ln\left(\frac{R_{n+1}-R_n}{R_{n+2}-R_{n+1}}\right).
\end{equation}

As with the detection of a shadow, the detection of a single photon ring has profound qualitative implications, requiring $N''(\rp)$ to be positive.\footnote{More properly, positive in an appropriate neighborhood of $\rp$, the size of which depends on the particular order photon ring under consideration.}  Otherwise, the photon orbit would be stable, and the associated null geodesics not reach distant observers to generate a ringlike structure in the images.

\begin{figure*}
    \centering
    \includegraphics[width=\columnwidth]{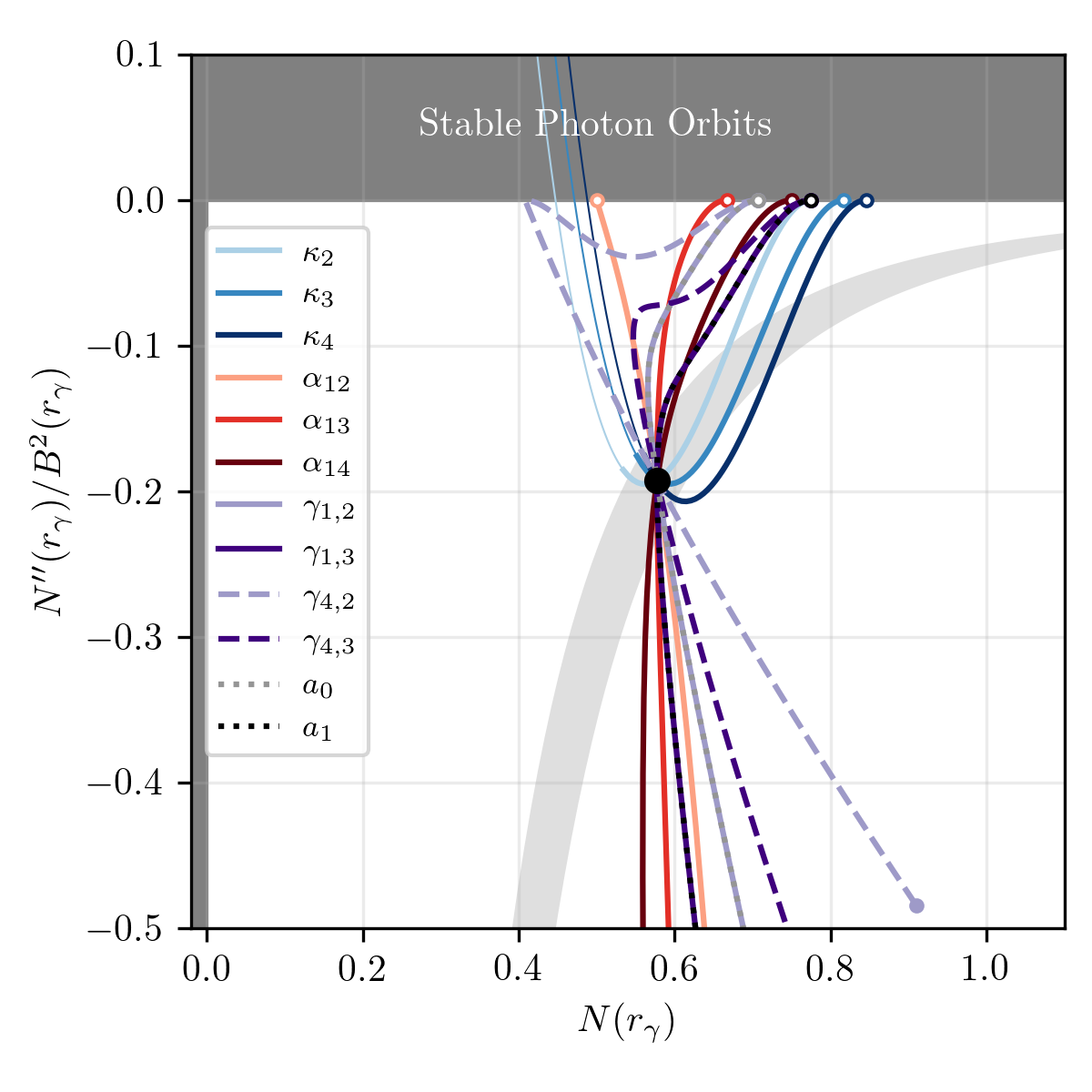}
    \includegraphics[width=\columnwidth]{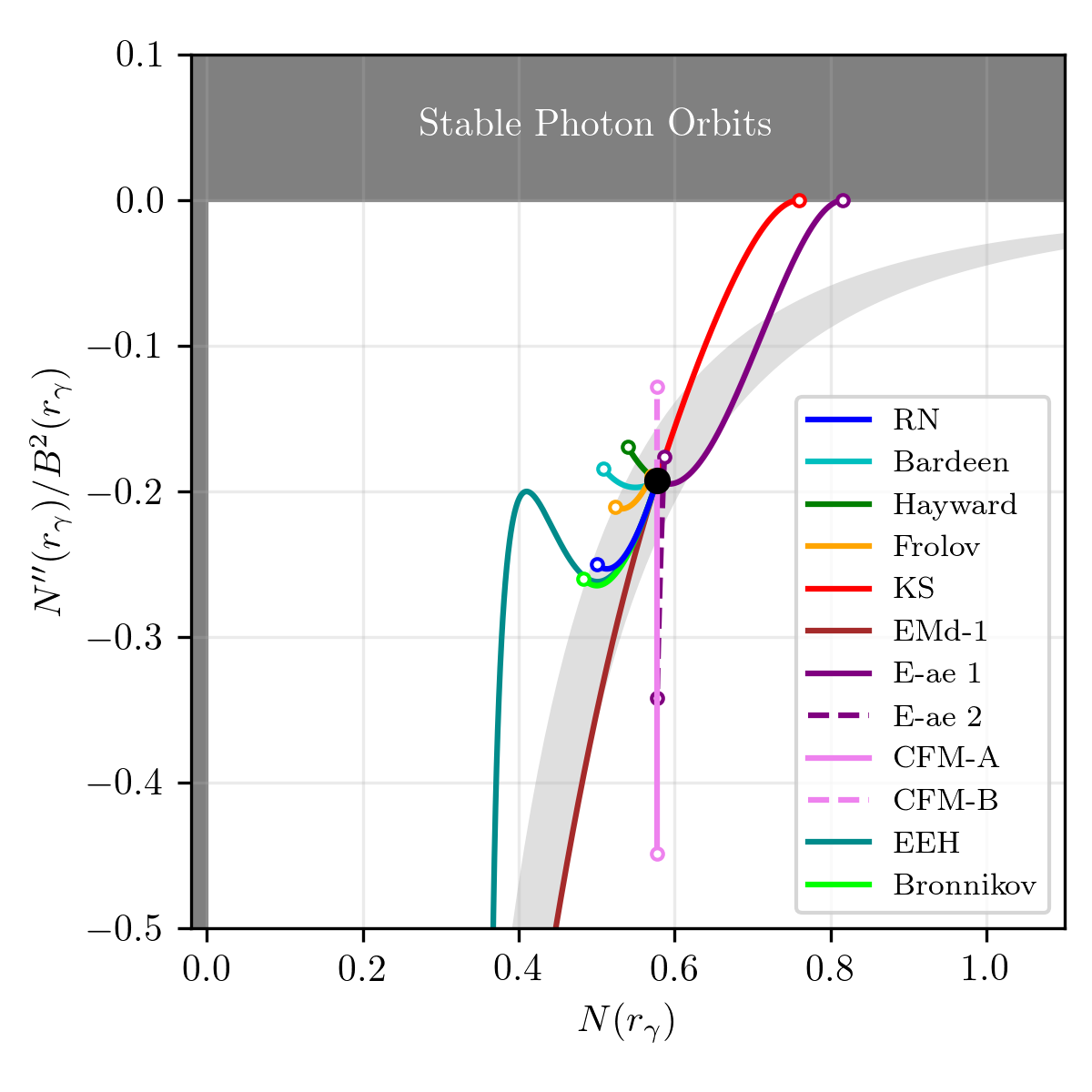}
    \caption{Constraints on $N''(\rp)/B^2(\rp)$ arising from a 10\% measurement of $\gamma$.  For reference, the curves associated with parameterized metrics (left) and alternative theories (right) are shown.  Line types and points are the same as in \autoref{fig:2pn} (left) and \autoref{fig:alts} (right).}
    \label{fig:lyapunovlimits}
\end{figure*}

Examples of the implications of a $10\%$ measurement of $\gamma$ are shown in \autoref{fig:lyapunovlimits}, roughly the precision required to distinguish the $n=1$ and $n=2$ photon rings.  Because measurements of $\gamma$ are necessarily coupled with a high-precision measurement of $R$, $N'(\rp)$ is effectively fixed, resulting in an additional constraint in the $N(\rp)$-$N''(\rp)/B^2(\rp)$ plane.  The qualitative differences between the interpretation of such a measurement for parameterized metric expansions and explicit metric alternatives is similar to that for the shadow size.  Again, the measurement presents a fundamentally degenerate constraint.  Thus, even multiple photon ring measurements, it is not possible to uniquely determine $N(\rp)$.

\subsection{Gravitational Waves}
With the detection of GW150914, gravitational waves have become an important probe of the near-horizon region of merging stellar-mass black holes.  Future space-based interferometers \citep[e.g., LISA][]{LISA2017} and pulsar timing monitoring experiments promise to expand these tests to supermassive black holes that are directly comparable to EHT observations of Sgr A* and M87*, respectively\citep{Hobbs2010}.

Comparison of these limits is complicated by the fact that gravitational wave observations necessarily require the specification of the dynamical sector of any putative gravity theory.  Therefore, it is generally insufficient to postulate alternative stationary spacetimes, as we have done in \autoref{eq:metric}, as doing so does not inform dynamical phenomena of any alternative theory. With this caveat, there are two situations in which may nevertheless be useful to characterize results in the way proposed here: gravitational wave ringdowns and extreme mass ratio inspirals (EMRIs), which we treat in turn.

\subsubsection{Gravitational Wave Ringdown}
The late-time evolution of high-angular momentum quasinormal modes (QNMs) is related in general relativity to the shadow size \citep{Jusufi2020,Ivan2020,Huan2021}.  This is a consequence of these modes being associated with high-frequency, azimuthally propagating massless perturbation, which is necessarily governed by the same dynamics as photons near the photon orbit.  As a consequence, generically the angular frequency, $\omega=\omega_{R,l} + i \omega_{I,l}$ of the quasinormal mode with azimuthal quantum number $l$,
\begin{equation}
    \lim_{l\rightarrow\infty} \omega_{I,l}
    =
    \frac{N'(\rp) \gamma}{2\pi}
    ~~\text{and}~~
    \lim_{l\rightarrow\infty} \frac{\omega_{R,l}}{l}
    =
    N'(\rp),
\end{equation}
which are equivalent to Equation I.1 of \citet{Huan2021}.  Thus, observations of high-frequency quasinormal modes result in spacetime constraints that are directly comparable to those from shadow sizes and multiple photon ring measurements.  Of particular interest is that these can be combined to separately measure $\gamma$:
\begin{equation}
    \gamma = 2\pi \lim_{l\rightarrow\infty} l \frac{\omega_{I,l}}{\omega_{R,l}}.
\end{equation}

While the above expressions invoke the high-$l$ limit, in practice, for Schwarzschild the approximations are good to better than 10\% by $l=2$, improving rapidly thereafter \citep[see, e.g.,][]{Berti2009}.  Of course, this does not confer any guarantees for alternative spacetimes.  Nevertheless, for illustrative purposes we consider the implications of the LIGO QNM measurements.\footnote{The LIGO sources are decidedly not spin-zero black holes.  The $\omega_R$ differ by $\sim30\%$ from their Schwarzschild counterparts for the typical merger remnant spins.  We defer a full discussion of the spinning case until \salehi.}

\begin{figure*}
    \centering
    \includegraphics[width=\columnwidth]{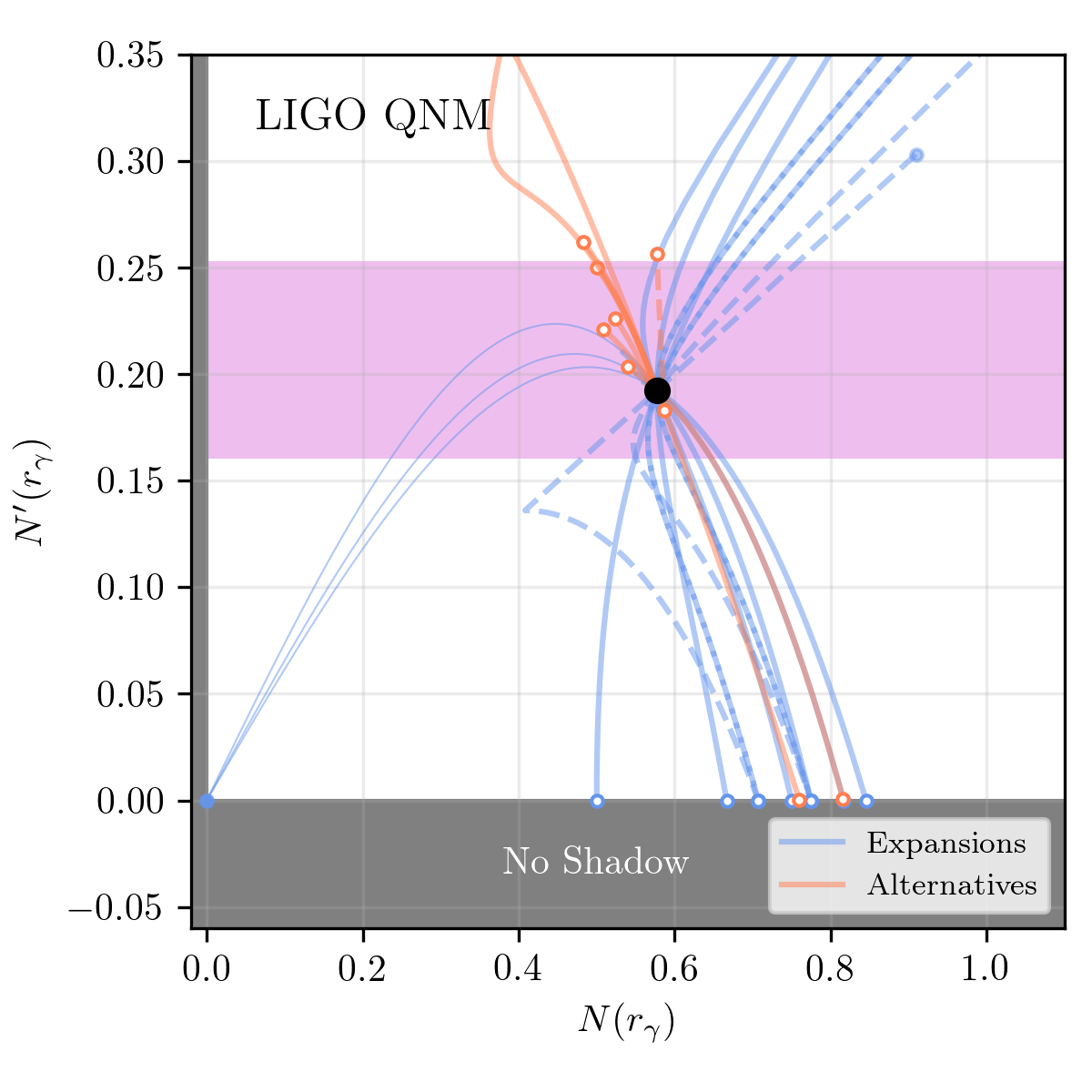}
    \includegraphics[width=\columnwidth]{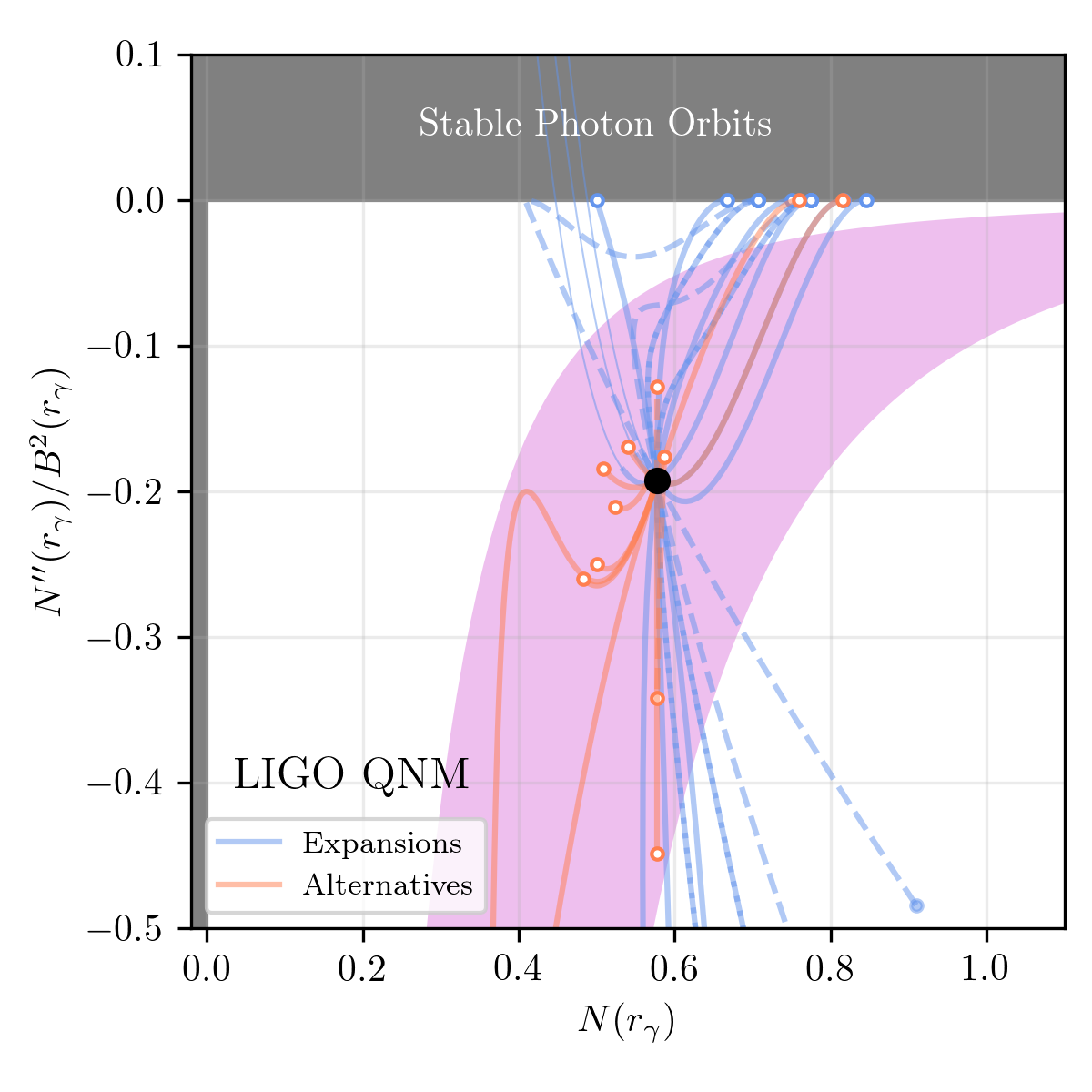}
    \caption{Constraints on $N(\rp)$, $N'(\rp)$, and $N''(\rp)/B^2(\rp)$ arising from the LIGO QNM constraints reported in \citet{LIGOQNM}.  The magenta shaded region shows the 68\% confidence region.  Illustrative curves associated with the metric expansions in \autoref{fig:2pn} (blue) and alternative spacetimes in \autoref{fig:alts} (orange) are shown.  See \autoref{fig:2pn} for a description of line types.}\label{fig:ligolimits}
\end{figure*}

\citet{LIGOQNM} reports constraints from LIGO on the low-order QNMs for 19 merger events.  For each event, analysis of the ringdown portion of the post-merger gravitational waveform yields an estimate of the QNM frequency ($f=\omega_{R,2}/2\pi$) and decay timescale ($\tau=1/\omega_{I,2}$).  Comparing to the expected values given the full inspiral-merger-ringdown waveform analyses finds that in aggregate the fractional shifts in these quantities are $\delta\hat{f}=0.03_{-0.21}^{+0.23}$ and $\delta\hat{\tau}=0.16\pm0.60$, respectively, where we have converted the 90\% region to $1\sigma$ assuming Gaussian statistics.  The implied aggregate measurement on $R$ is
\begin{equation}
    R = \frac{\sqrt{27}}{1+\delta\hat{f}}
    = \sqrt{27}(0.97_{-0.18}^{+0.25}),
\end{equation}
and the implied $\gamma$ is, then,
\begin{equation}
    \gamma = \frac{\pi}{(1+\delta\hat{f})(1+\delta\hat{\tau})}
    =
    1.9_{-0.3}^{+3.3}.
\end{equation}
Note that the uncertainties on $R$ and $\gamma$ are correlated through $\delta\hat{f}$.  The corresponding limits on $N(\rp)$, $N'(\rp)$, and $N''(\rp)$ are shown in \autoref{fig:ligolimits} in comparison to various parameterized metric expansions and alternative metrics.

\subsubsection{Gravitational Wave Inspiral}
\label{sec:emri}
EMRIs have the significant virtue of the secondary being a perturbation on the spacetime of the much more massive primary \citep[see, e.g.,][and references therein]{LISA_EMRIs}. Therefore, as with QMNs, the dynamics of the EMRI system can be analyzed within the context of a stationary background spacetime, subject to the caveat regarding the dynamical sector of the underlying gravity theory.  However, there is an additional complication: the inspiral typically occurs at $r>\rp$.  

A short discussion of the circular orbits of massive paricles in arbitrary spherically symmetric spacetimes may be found in \autoref{app:orbits}.  An implicit expression for the radius of the ISCO, which terminates the inspiral regime, is given in \autoref{eq:risco}.  While it is clear that the detection of a photon ring ensures that stable circular orbits do not exist a $\rp$ (see \autoref{app:orbits}), general statements about the ordering and proxmity of ISCO and photon orbit depend on the particular form of $N(r)$.  As a result, some assumption regarding the extrapolation of $N(r)$ away from $\rp$ is necessary to make any estimate of the implications of inspiral waveform measurements for deviations from general relativity.

\section{Conclusions}
\label{sec:conc}
The EHT images of M87* and Sgr A* provide a new, direct window into the properties of astrophysical black holes.  Shadow size measurements, and more generally, measurements of the size of photon rings, provide a means to directly probe the spacetime geometry of black holes.  However, interpreting the gravitational implications of these measurements requires some care due to the nonlinear nature of general relativity near black hole event horizons and limited information contained in a single (or handful) of size measurements.  This is evident even in the limited case of spherically symmetric, stationary spacetimes that we address.

For such spacetimes, significant qualitative conclusions may be reached already based solely upon the detection of various image features.  The detection of a shadow immediately implies that $N'(\rp)>0$.  The detection of any photon ring implies that $N''(\rp)<0$. These hold independent of the shadow or photon ring size.

The size of the shadow is directly related to $N'(\rp)$: measuring the shadow size is synonymous with measuring the radial derivative of the $tt$-component of the metric in areal coordinates.  In this sense, precise shadow size measurements generate precise metric constraints.  However, there is no constraint on $N(\rp)$ from the shadow size alone, and thus shadow size measurements by themselves do not provide any limit on the magnitude of a putative deviation in the $tt$-component of the metric, i.e., the perturbing potential $\psi(r)$.  In this sense, precise shadow measurements are uninformative.

When translated to spacetime parameters, either via parameterized metric expansions or explicit alternative spacetimes, precise shadow size measurements can appear to impose strong constraints on both $N(\rp)$ and $N'(\rp)$.  The constraint on $N(\rp)$ is a direct consequence of the prior that has been adopted via the choice of the underlying metric.  In the case of parameterized metric expansions, for which there is little significance to this prior, the strength of the attendant limit on $N(\rp)$ is illusory.  Indeed, it is straightforward to generate examples of perturbed metrics that are otherwise consistent with all existing constraints for Sgr A* and M87*, including the recent EHT shadow sizes, that have nearly arbitrary magnitude metric perturbations at the photon orbit.  Claims in the literature that shadow size measurements limit a specific metric expansion coefficient, or a linear combination of coefficients, must be understood within a narrow context for a similar reason.

Nevertheless, because shadow size measurements do impose a strong measurement on some facet of the underlying spacetime, we propose an alternative way to characterize their gravitational implications: the values of $N(r)$, $N'(r)$, $N''(r)/B^2(r)$, etc., measured at $\rp$.  This series has the virtue of being a nonparametric and nonperturbative description of the shadow size measurement --- in spherical symmetry the identifications are exact.  However, they are complicated by the a priori unknown value of $\rp$, i.e., properties of the metric are constrained at an important dynamical location in the spacetime whose location is otherwise unknown.

Despite the unknown $\rp$, this proposal presents a particularly useful basis for comparing near-horizon measurements, including redshifts, photon echos, photon rings, shadow sizes, and black hole ringdowns, all of which are dominated by massless particle dynamics near the event horizon, and therefore the photon orbit.  It is also convenient theoretically, providing a more natural quantity for theoretical comparison and obviating the need for full ray-tracing and radiative transfer simulations and/or mode spectrum computation. 

This basis remains poorly connected to gravity measurements that probe very different spatial scales.  This is a natural consequence of the afore-mentioned nonlinearity expected near the photon orbit: the unavoidable price of adopting a nonperturbative way to characterize near-horizon observations is difficulty in making comparisons to perturbative characterizations of other measurements.  This complicates, e.g., quantitatively relating EHT shadow size measurements to solar system tests or observations of the inspiral phase of black hole mergers outside of a particular gravity theory.  

We have focused on spherically symmetric spacetimes for simplicity.  However, to linear order in spin, all of the results obtained for Schwarzschild continue to apply for polar observers (the relevant inclination for M87*). We leave the expansion of the nonparametric, nonperturbative characterization to integrable rotating spacetimes with arbitrary spins for future work.  Nevertheless, even the simple cases explored here, elucidate the power and limitations of measurements of the shadow and photon ring sizes.

\begin{acknowledgments}
This work was supported in part by Perimeter Institute for Theoretical Physics.  Research at Perimeter Institute is supported by the Government of Canada through the Department of Innovation, Science and Economic Development Canada and by the Province of Ontario through the Ministry of Economic Development, Job Creation and Trade.  A.E.B. thanks the Delaney Family for their generous financial support via the Delaney Family John A. Wheeler Chair at Perimeter Institute.  A.E.B. receives additional financial support from the Natural Sciences and Engineering Research Council of Canada through a Discovery Grant.
\end{acknowledgments}

\appendix

\section{Slowly Rotating Space times}
\label{app:slowrot}
When viewed from the polar axis, approximately appropriate for M87*, by symmetry all axisymmetric spacetimes must produce circular shadows.  While limiting ourselves to this viewing angle reduces the impact of spacetime properties, it does simplify the relationship between the shadow size and the underlying spacetime geometry.

In \salehi, the shadows of a broad class of integrable, asymptotically flat, axisymmetric, spinning spacetimes is studied in detail.  In additional, these spacetimes reduce to spherical symmetry in the limit of zero spin, i.e., $a=0$.  Each may be written in the form
\begin{equation}
\begin{aligned}
    ds^2 
    =& 
    -N(r)^2\left[1 + a^2 f_t(r,\cos\theta;a^2)\right] dt^2 \\
    &+\frac{B(r)^2}{N(r)^2}\left[ 1 + a^2 f_r(r,\cos\theta;a^2)\right] dr^2 \\
    &+r^2\left[1 + a^2 f_\theta(r,\cos\theta;a^2)\right] d\theta^2 \\
    &+r^2\sin^2\theta\left[1 + a^2 f_\phi(r,\cos\theta;a^2)\right] d\phi^2 \\
    &+ 2 a g(r,\cos\theta;a^2) dt d\phi,
\end{aligned}
\end{equation}
for some set of functions $f_t(r,\cos\theta;a^2)$, $f_r(r,\cos\theta;a^2)$, $f_\theta(r,\cos\theta;a^2)$, $f_\phi(r,\cos\theta;a^2)$, and $g(r,\cos\theta;a^2)$, 
where we have explicitly enforced the reduction to spherical symmetry when $a=0$ and reflection symmetry across the equatorial plane.

Because of this latter symmetry, shadow sizes observed by polar viewers must be independent of the sign of the spin.  Expanding the above confirms that the lowest-order spin correction enters at $a^2$ \salehi.  As a consequence, up to corrections of $a^2$, the shadow size is that given in \autoref{eq:Rdef}.

\section{Explicit Metric Expansions}
\label{app:expexp}
Here we collect explicit expansions for the various metric expansions.  In all cases, to produce the single-parameter curves in \autoref{fig:2pn} it is easier to solve for the coefficient as a function of $\rp$.  Care must be taken when more than one photon orbit exists (which occurs for the MGBK metric when only $\gamma_{4,2}$ is varied).  All of the expansions described here are similar in character and reduce to expansions in $1/r$ for small parameter values.  We adopt the form of the metric expansions used in \citet{SgpaperVI}.

\subsection{Post-Newtonian}
Combining \autoref{eq:N2psi} and \autoref{eq:PNpsi} gives the expression for $N^2(r)$ for the post-Newtonian metric expansion:
\begin{equation}
    N^2(r) = 1 - \frac{2M}{r} + \sum_{n=1}^\infty \frac{2\kappa_n (-M)^{n+1}}{r^{n+1}}.
\end{equation}

\subsection{JP Metric}
From \citet{Johannsen2013},
\begin{equation}
    N^2(r) = \left(1-\frac{2M}{r}\right)
    \left(1 + \sum_{n=2}^\infty \frac{\alpha_{1n} M^n}{r^n}\right)^{-2}.
\end{equation}

\subsection{MGBK Metric}
From \cite{Vigeland2011}, 
\begin{multline}
    N^2(r) = \left(1-\frac{2M}{r}\right)\Bigg[
    1 - \sum_{n=2}^\infty \frac{\gamma_{1,n} M^n}{r^n}\\
    - 2 \left(1-\frac{2M}{r}\right) \sum_{n=2}^\infty \frac{\gamma_{4,n} M^n}{r^n} \Bigg].
\end{multline}
When only varying $\gamma_{4,2}$, this metric admits up to three solutions for $\rp$.  Where appropriate, the largest $\rp$ is adopted, which is evident in \autoref{fig:2pn}.

\subsection{RZ Metric}
From \citet{RZ14}, up to $a_1$,
\begin{equation}
    N^2(r) = \left(1-\frac{r_0}{r} \right)
    \left[ 1 - \epsilon \frac{r_0}{r} 
    + (a_0-\epsilon) \frac{r_0^2}{r}^2
    + a_1 \frac{r_0^3}{r^3}
    \right].
\end{equation}
Imposing $r_0=2M$ and $\epsilon=-(1-2M/r_0) = 0$, as done in \citet{SgpaperVI}, reduces this to
\begin{equation}
    N^2(r) = \left(1-\frac{2M}{r} \right)
    \left[ 1
    + 4a_0 \frac{M^2}{r^2}
    + 8a_1 \frac{M^3}{r^3}
    \right],
\end{equation}
which is degenerate with the MGBK metric expansion with $\gamma_{1,2} = 4a_0$ and $\gamma_{1,3} = 8a_1$.

\section{Shadows and Photon Orbits}
\label{app:shadows}
We review some facts about the shadow and their relationship to the photon orbit in arbitrary spacetimes.  We begin with the equation of motion for a null geodesic in an arbitrary stationary and spherically symmetric spacetime, which we obtain from the null condition and the fact that $E$ and $L=Eb$ are conserved:
\begin{equation}
    \dot{r}^2 = \frac{1}{B^2(r)} \left[ 1 - \frac{b^2}{r^2} N^2(r) \right],
    \label{eq:eom}
\end{equation}
where $\dot{r}=dr/d\tau$.  An effective potential, $V_{\rm eff}(r)=N^2(r)/r^2$, may be specified in the normal way, yielding
\begin{equation}
    \dot{r}^2 = \frac{1}{B^2(r)} \left[ 1 - b^2 V_{\rm eff}(r) \right].
\end{equation}
Turning points occur when $V_{\rm eff}(r) = 1/b^2$.  If $V_{\rm eff}(r)$ has a maximum, there will be a $b$ for which all null geodesics with smaller $b$ will have no turning points between the event horizon and the observer at infinity.  The collection of such trajectories comprise the shadow.

At a photon orbit, two conditions apply: $\dot{r}=0$ (i.e., the orbit is circular) and $\ddot{r}=0$ (i.e., it stays circular), and thus
\begin{equation}
    b_\gamma = \frac{1}{\sqrt{V_{\rm eff}(\rp)}} = \frac{\rp}{N(\rp)}
    \label{eq:bgamma}
\end{equation}
and
\begin{equation}    
    \left[\frac{1}{V_{\rm eff}(r)}\right]''_{\rp} = 
    \left[ \frac{r^2}{N^2(r)} \right]'_{\rp} = 0,
    \label{eq:photon_orbit}
\end{equation}
which are equivalent to \autoref{eq:Rdef} and \autoref{eq:rpdef}, respectively.  

Note that these conditions may be satisfied at more than one location, corresponding to multiple photon orbits.  That there is at least one unstable photon orbit for asymptotically flat, black hole spacetimes is guaranteed by two facts: 1.\ asymptotic flatness gives $V_{\rm eff}(r)>0$ and $V'_{\rm eff}(r)<0$ at very large $r$, and 2.\ $V_{\rm eff}(\rh)=0$ vanishes at some $\rh$ that defines the event horizon.  Thus, $V_{\rm eff}$ is positive and decreasing at large $r$ and vanishes at small $r$, requiring via the mean value theorem that at some intermediate point $V_{\rm eff}(r)$ has a maximum.  Subsequently, photon orbits may appear in pairs, one stable (maximum of $V_{\rm eff}(r)$) and one unstable (minima of $V_{\rm eff}(r)$).  Because the shadow is necessarily associated with the global maximum of $V_{\rm eff}(r)$, by \autoref{eq:bgamma}, its size is set by the smallest $b_\gamma$.  Thus, henceforth, we will restrict ourselves to the photon orbit with the smallest shadow, which we call $\rp$ and $b_\gamma$, respectively.

\section{Radii of Photon Rings}
\label{app:photonrings}
The higher order images of the near-horizon emission region form a sequence of typically ring-like image features \citep[see, e.g.]{BroderickLoeb2006,Johnson2020,Spin}.  Each additional image is associated with photon trajectories that execute an additional half orbit about the black hole, and therefore the size of these rings are probes of spacetime about the photon orbit.  Here we derive the radius of the higher-order photon rings in the limit of high photon-ring order.

For null geodesics near the photon orbit, i.e., for which 
\begin{equation}
    r=\rp+\delta r,
\end{equation}
\autoref{eq:eom} becomes after Taylor expanding about $r=\rp$ to lowest order,
\begin{equation}
    \dot{\delta r}^2 \approx - \frac{b_\gamma^2}{2B^2(\rp)} \left[\frac{N^2(r)}{r^2}\right]''_{\rp} \delta r^2,
\end{equation}
where the first two terms vanish due to \autoref{eq:bgamma} and \autoref{eq:photon_orbit} and because the inner turning point (i.e., the minimum $\delta r$) is small, we've replaced $b$ with $b_\gamma$.  Combining this radial equation of motion with \autoref{eq:consts}, gives
\begin{equation}
    \pi \frac{d \delta r}{d\phi}
    =
    \gamma\delta r
\end{equation}
where
\begin{equation}
    \gamma = \pi \frac{N^{3/2}(\rp)}{N'(\rp)} 
    \left[ -\frac{N''(\rp)}{B^2(\rp)} \right]^{1/2}
\end{equation}
is the Lyapunov exponent identified in \citet{Johnson2020}.  Solving this, we obtain,
\begin{equation}
    \delta r = \delta r_0 e^{\gamma \phi/\pi},
\end{equation}
i.e., upon every half orbit the geodesic grows by a factor of $e^\gamma$.  This matches Equation 7 of \citet{Johnson2020} in the limit of Schwarzschild.

At some $\delta r_{\rm max} \sim M$, the perturbative expansion of the radial equation of motion breaks down, and the photon trajectory streams to the observer at infinity.  Prior to this point, it will have executed
\begin{equation}
    n \approx 2 \gamma^{-1} \ln\left(\delta r_{\rm max}/\delta r_0\right),
\end{equation}
half orbits, where the preceding orbits from a distant source to the vicinity of the photon orbit are now included, and thus contribute to the $n$th-order photon ring.  Describing the details of the transition is not necessary to obtain the relative locations of the photons on a distant observing screen; rather the photon ring sizes are completely controlled by $b$, which is, in turn, set by the inner turning point, i.e., by $\delta r_0$.  At $\delta r_0$, $\dot{r}=0$, and therefore,
\begin{equation}
    b^2 \approx b^2_\gamma + \frac{1}{2} \left[ \frac{r^2}{N^2(r)} \right]''_{\rp} \delta r_0^2,
\end{equation}
where again the linear term vanishes as a result of \autoref{eq:photon_orbit}.  Identifying the radius of the $n$th order photon ring, $R_n$, with the $b$ associated with the $\delta r_0$ that corresponds to $n$ half orbits, we find
\begin{equation}
    R_n-R \approx
    \frac{N'(\rp)}{4} \left[ \frac{r^2}{N^2(r)} \right]''_{\rp} \delta r_{\rm max}^2 e^{-\gamma n}.
\end{equation}
The prefactor is common to all order photon rings.  The shift relative to the shadow size decreases as $e^{-\gamma n}$, from which we recover that the shadow is bounded by the asymptotic photon ring corresponding to $n\rightarrow\infty$.  This matches Equation 12 of \citet{Johnson2020}.

\section{Circular Orbits of Massive Particles}
\label{app:orbits}
General expressions for circular orbits for massive particles may be constructed for the family of spacetimes described by \autoref{eq:metric} in a fashion similar to that used to obtain $\rp$.  We again require $\dot{r}$ and $\ddot{r}$ vanish. The first condition gives
\begin{equation}
    \dot{r}^2 = \frac{1}{B^2(r)} \left[E^2 - N^2 - \frac{N^2}{r^2} L^2 \right] = 0,
\end{equation}
where $E=-u_t$ and $L=u_\phi$ are the conserved specific energy and specific angular momentum.  From this, the second gives
\begin{equation}
    \ddot{r}
    =
    \frac{N(r)}{B^2(r)}
    \left[
    N'(r) + L^2 \left(\frac{N'(r)}{r^2}-\frac{N(r)}{r^3}\right) 
    \right]
    = 0, 
\end{equation}
where we employed the previous condition.  Together, these imply that for circular orbits,
\begin{equation}
    L\equiv u_\phi = \pm\sqrt{\frac{r^3 N'(r)}{N(r)-rN'(r)}},
\end{equation}
with associated energy,
\begin{equation}
    E\equiv u_t = -\sqrt{\frac{N^3}{N(r)-rN'(r)}}.
\end{equation}
From these, the angular velocity as measured by a distant observer is,
\begin{equation}
    \Omega = \frac{u^\phi}{u^t} = -\frac{N^2(r)}{r^2}\frac{L}{E} = \pm \sqrt{\frac{N(r)N'(r)}{r}}.
    \label{eq:Omega}
\end{equation}

The stability of these circular orbits is determined by the response to perturbations (again similar to the analysis of photon orbits in \autoref{app:photonrings}),
\begin{equation}
    \dot{\delta r}^2 = -\omega^2 \delta r^2
\end{equation}
where
\begin{equation}
    \omega^2 = \frac{N(r)}{B^2(r)} \left[
        N''(r) \left(1+\frac{L^2}{r^2}\right)
        +
        \frac{3N^3(r)}{r^4} \frac{L^2}{E^2}
    \right] \delta r^2.
\end{equation}
When $\omega^2>0$, the perturbation is oscillatory and the orbits are stable; when $\omega^2<0$ the perturbation grows exponentially.  Note that at $\rp$, $L^2\rightarrow\infty$ and $\omega^2<0$ if $N''(\rp)<0$, i.e., if any photon rings are observed then timelike geodesics are also unstable at the photon orbit.  

Transitions from stable to unstable circular orbits occur when $\omega^2=0$.  The radius of the innermost stable circular orbit is the minimum $\risco$ for which
\begin{equation}
    \risco = - \frac{3N(\risco)N'(\risco)}{3N'^2(\risco)-N(\risco)N''(\risco)},
    \label{eq:risco}
\end{equation}
which is an analogous condition to that for $\rp$ in \autoref{eq:rpdef}.  However, in the absence of knowledge about the particular form of $N(r)$ away from $\rp$, it is difficult to place any further general conditions on $\risco$ relative to $\rp$.

\clearpage
\bibliographystyle{aasjournal}
\bibliography{references.bib}

\end{document}